\documentclass[cameraready]{Interspeech}

\title{Efficient Emotion and Speaker Adaptation in LLM-Based TTS via Characteristic-Specific Partial Fine-Tuning}

\author[affiliation={1}]{Tianrui}{Wang}
\author[affiliation={1}]{Meng}{Ge}
\author[affiliation={2}]{Cheng}{Gong}
\author[affiliation={1}]{Chunyu}{Qiang}
\author[affiliation={1}]{Haoyu}{Wang}
\author[affiliation={1}]{\\Zikang}{Huang}
\author[affiliation={1}]{Yu}{Jiang}
\author[affiliation={3}]{Ye}{Ni}
\author[affiliation={1}]{Yuheng}{Lu}
\author[affiliation={1}]{Xiaobao}{Wang}
\author[affiliation={6}]{\\Engsiong}{Chng}
\author[affiliation={4}]{Xie}{Chen}
\author[affiliation={1,5}, correspondingauthor]{Longbiao}{Wang}
\author[affiliation={7}]{Jianwu}{Dang}

\address{
  $^1$ Tianjin Key Laboratory of Cognitive Computing and Application, College of Intelligence and Computing, Tianjin University, Tianjin, China 
  $^2$ TeleAI, China Telecom, Beijing, China \\
  $^3$ Southeast University, Nanjing, China 
  $^4$ Shanghai Jiao Tong University, Shanghai, China \\
  $^5$ Huiyan Technology Co., Ltd, Tianjin, China 
  $^6$ Nanyang Technological University, Singapore \\
  $^7$ Shenzhen Institute of Advanced Technology, Chinese Academy of Sciences, Guangdong, China 
}

\email{\{wangtianrui, gemeng, gongchengcheng, qiangchunyu, hy\_wang, huangzikang, jiang\_y, luyuheng2024, wangxiaobao, longbiao\_wang\}@tju.edu.cn, niye@seu.edu.cn, aseschng@ntu.edu.sg, chenxie95@sjtu.edu.cn, jdang@jaist.ac.jp}


\usepackage{comment}
\usepackage{amssymb}
\usepackage{tabularx}
\usepackage{makecell}
\usepackage{rotating}
\usepackage{threeparttable}
\usepackage{booktabs}
\usepackage{array}
\usepackage{multirow}
\usepackage{subfigure}
\usepackage{verbatim}
\usepackage{lineno,hyperref}
\usepackage{amsmath,bm}
\newcommand\our{\textsc{CSP-FT}}

\begin{document}

\maketitle

\begin{abstract}
While LLM-based TTS models exhibit zero-shot emotion and speaker cloning, their cloning fidelity and pronunciation clarity degrade on unseen domains. Fine-tuning is essential for adaptation, yet uniform approaches overlook specific parameter contributions. Uniform tuning on limited data causes slow training and catastrophic forgetting, leading to degraded pronunciation accuracy. To address this, we propose \textbf{\our{}}, a \textbf{c}haracteristic-\textbf{s}pecific \textbf{p}artial \textbf{f}ine-\textbf{t}uning strategy. 
By dynamically analyzing layer contributions via a weighted sum, we selectively fine-tune only the two layers capturing the most and least emotion and speaker information, maximizing the utility of the former while explicitly strengthening the capacity of the latter.
Experiments on a combined corpus of 11 datasets show \our{} matches or exceeds the fidelity and intelligibility of full fine-tuning while updating only $\sim$8\% of parameters, accelerating training by $\sim$2$\times$, and significantly mitigating catastrophic forgetting.
\end{abstract}

\section{Introduction}
\label{intro}
With advancements in text-to-speech (TTS) technology, the focus has expanded beyond achieving high content intelligibility to include emotional expression and speaker cloning \cite{triantafyllopoulos2023overview, tts1}. To improve task support and model performance, TTS models are being developed with increasingly larger parameters and trained on larger datasets \cite{tts2, tts3}. As many large-scale general codec language TTS models are now open-sourced, adapting them to generate speech that meets specific emotional expression and speaker cloning requirements has become a growing area of research interest \cite{intro_ft3, neekhara2021adapting}.

Adapting general pre-trained TTS models to generate speech with specific emotional expression and speaker identity is fundamentally a domain adaptation (DA) problem \cite{singhal2023domain}. Based on the availability of target domain data, DA methods are generally categorized into data-scarce, task-irrelevant, and target-data-available approaches. Data-scarce DA (e.g., few-shot) focuses on adapting a pre-trained model using limited examples, typically through in-context learning or fine-tuning \cite{choi2020attentron, huang2022meta}. Meanwhile, task-irrelevant DA (e.g., zero-shot) relies on the model's inherent generalization capabilities to adapt to new tasks without fine-tuning \cite{xin2024rall, casanova2022yourtts, valle}. While open-sourced codec language models demonstrate zero-shot capabilities for prompted emotion and speaker identity, their performance often remains unstable, particularly with unseen speakers or emotional expressions \cite{vallex, singhal2023domain}. In contrast, target-data-available DA employs supervised fine-tuning with a small amount of labeled target data. By explicitly aligning the model with the target domain, this approach significantly improves the control of emotion and speaker identity, delivering much more stable and reliable performance \cite{motiian2017unified, neekhara2021adapting, huang2024voicetuner, lou2024stylespeech}.

\begin{figure}[t]
\centering
\includegraphics[width=8cm]{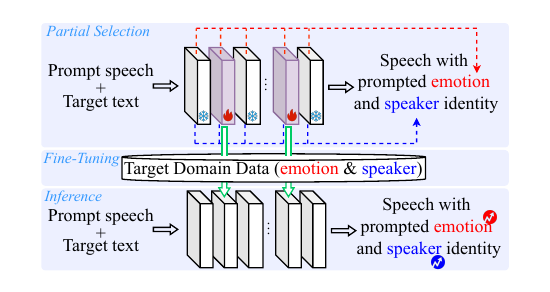}
\vspace{-0.5cm}
\caption{
Characteristic-specific partial fine-tuning for emotion and speaker adaptation in codec language TTS models.
} 
\vspace{-0.5cm}
\label{intro_pic}
\end{figure}

\begin{figure*}[t]
\centering
\includegraphics[width=17cm]{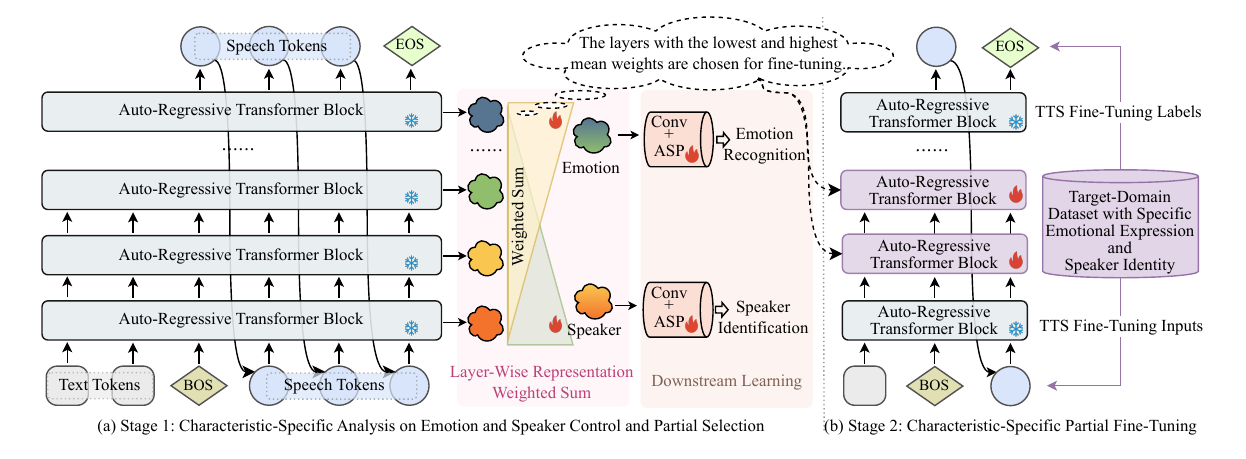}
\vspace{-0.5cm}
\caption{Overview of the proposed approach. The codec language TTS model is fine-tuned for emotion recognition and speaker identification tasks using a weighted-sum framework. The method prioritizes layers with the lowest and highest weighted importance, facilitating efficient fine-tuning to generate speech with target-domain specific speaker and emotional expressions.
} 
\label{proposed}
\end{figure*}

However, effective target-data-available DA must balance the learning of target knowledge with the critical challenges of resource consumption and catastrophic forgetting. Fine-tuning methods are broadly categorized into full fine-tuning \cite{intro_ft, intro_ft2, superb, intro_ft3} and parameter-efficient fine-tuning (PEFT) \cite{hu2022lora, adapter, wang2023adapter}. Full fine-tuning adapts the entire model’s parameters to the target task, which demands substantial computational resources and is prone to overfitting the limited target data. This overfitting leads to catastrophic forgetting, where the model loses significant pre-trained foundational knowledge \cite{forgetting, mccloskey1989catastrophic, french1999catastrophic, goodfellow2013empirical}. This degradation severely compromises the model's ability to generalize to fine-tuning-unseen data; for example, in TTS tasks, fine-tuning a model for a specific speaker can impair other critical capabilities such as word accuracy and general emotion control \cite{fan2015multi}. Conversely, PEFT freezes all pre-trained parameters and only trains lightweight inserted modules to minimize resource consumption, but this often ignores the specific contributions of the original network's parameters.

To mitigate these limitations, we propose characteristic-specific partial fine-tuning, short as \our{}. As illustrated in Figure~\ref{intro_pic}, our method partially adapts a pre-trained codec language TTS model using a small amount of target-domain data by evaluating the varying contributions of different Transformer layers. Specifically, we employ a weighted-sum approach to analyze the contributions of each layer to emotion recognition and speaker identification tasks. To simultaneously adapt to target-domain characteristics, we selectively fine-tune the layer with the highest contribution (to leverage its maximal control) alongside the layer with the lowest contribution (which holds the greatest potential for improvement).
Our main contributions are summarized as follows:
\begin{itemize}
\item We propose \our{}, a novel partial fine-tuning strategy that explicitly targets specific Transformer layers based on their characteristic contributions to emotion and speaker control.
\item Experiments on four open-source models—GPT-SoVITS\footnote{\url{https://github.com/RVC-Boss/GPT-SoVITS}}, VALLE-X \cite{vallex}, CosyVoice \cite{cosyvoice}, and Fun-CosyVoice3.0 \cite{du2025cosyvoice3}—demonstrate that \our{} matches or surpasses full fine-tuning performance, achieving $2\times$ faster training speeds, minimal trainable parameters, and significantly reduced catastrophic forgetting.
\item We validate the strong cross-dataset robustness of our method, proving that the identified layer contribution profiles can be universally applied to new target-domain data without requiring repeated analysis.
\item We demonstrate that generative codec language TTS models can function as highly effective speech encoders for perception tasks (e.g., emotion recognition and speaker identification), offering valuable insights for universal speech processing.
\end{itemize}

\section{Methodology}
\label{method}
Our objective is to adapt pre-trained codec language TTS models to generate speech with target-domain emotional expressions and speaker identities. While full fine-tuning achieves this, it computationally slows the process and risks catastrophic forgetting of pre-trained content knowledge. To mitigate this, we propose \our, a partial parameter fine-tuning strategy based on the varying contributions of different model components to emotion and speaker control, as illustrated in Figure~\ref{proposed}.  

\subsection{Preliminary and Overview of the Proposed Method}
The goal of domain adaptation in TTS is to effectively transfer a pre-trained model $f: \mathbf{X}^s \to \mathbf{Y}^s$ from a source domain to a target domain $f: \mathbf{X}^t \to \mathbf{Y}^t$, enabling it to generate speech with target-specific emotional expressions and speaker identities. The adapted model $f$ must excel on the target domain while retaining the foundational knowledge acquired during pre-training. Unlike traditional approaches that treat all parameters equally, our characteristic-specific partial fine-tuning strategy selects a specialized subset of parameters based on their contributions to emotion and speaker control. By freezing the remaining parameters, we exclusively fine-tune this selected subset, which accelerates training and drastically reduces the risk of catastrophic forgetting.

As shown in Figure~\ref{proposed}, \our{} operates in two stages: characteristic-specific analysis and targeted partial fine-tuning. First, we repurpose the codec language TTS model as an encoder, utilizing two sets of learnable weights to compute weighted sums of the Transformer layer outputs. These representations are then fed into lightweight downstream modules to perform emotion recognition and speaker identification tasks. Second, we average the two sets of layer weights to identify the Transformer blocks with the highest and lowest contributions. Only these targeted layers are fine-tuned on the target domain. This explicit selection process ensures the preservation of pre-trained knowledge while maximizing the model's capacity for target-domain emotional and speaker expression.

\subsection{Characteristic-Specific Analysis of Codec Language TTS}
\label{superb_ft}
The codec language TTS model uses a Transformer architecture to predict the next discrete speech token in an auto-regressive manner based on text and speech prompts. We repurpose this pre-trained model as a causal encoder, fine-tuning it with a learnable weighted-sum strategy for emotion recognition and speaker identification, enabling an analysis of how the model's layers capture and control emotional and speaker-specific information in speech.

The input speech waveform is converted into discrete speech tokens $\mathbf{A} \in \mathbb{R}^{T_A \times 1}$ using a codec compression model with quantization \cite{encodec, cosyvoice}. Then, the text, beginning of sentence token (BOS), and speech tokens are passed through embedding modules to generate continuous representations, which are concatenated and fed into an $N$-layer Transformer. The Transformer processes the input and produces $D$-dimensional layer-wise representations, denoted as $\mathbf{O} \in \mathbb{R}^{N \times (T_S+1+T_A) \times D}$:
\begin{equation}
   \mathbf{O} = \left\{\bm{O}_1, \bm{O}_2, \dots, \bm{O}_N\right\}.
\end{equation}

In speech synthesis, the model directly predicts the next speech discrete token based on $\bm{O}_N$. However, our goal is to analyze the contribution of each layer of the Transformer to emotion and speaker control in the generated speech. To achieve this, we apply normalization and combine the layer outputs using learnable weights with softmax, $\bm{W}_e$ or $\bm{W}_s$, thereby obtaining task-specific representations $\bm{Z}_e, \bm{Z}_s \in \mathbb{R}^{(T_S+T_A+1) \times D}$, defined as:
\begin{equation} 
\bm{Z} = \sum_{i=1}^N\left(\frac{e^{\omega_i}}{\sum_{j=1}^N e^{\omega_j}} \cdot \text{layernorm}(\bm{O}_i)\right), \label{weight} 
\end{equation}
where $\bm{W}=\left\{\omega_1, \omega_2, \dots, \omega_N\right\}$ represents the learnable weights for each layer.

\label{downstream_models}
The contribution of controlling emotion and speaker identity in the generated speech is primarily reflected in the amount of relevant information contained by the module's representations. Therefore, we use speech emotion recognition and speaker identification tasks to train $\bm{W}$. Specifically, the characteristic-specific representations $\bm{Z}_e, \bm{Z}_s$ are transformed into utterance-level representations. This transformation is achieved using a stacked convolution module (Conv), consisting of 1D Convolution and ReLU, followed by attentive stat pooling (ASP) \cite{desplanques2020ecapa}, which aggregates the frame-level representations. Finally, a cross-entropy loss function is applied to optimize the layer weights, convolutions, and ASP modules. 

After training on speech emotion recognition and speaker identification, the layer weights $\bm{W}_e$ and $\bm{W}_s$ indicate the contribution of each layer for the respective downstream tasks. By analyzing the magnitude of these weights, we can identify the relevant layers for fine-tuning in order to generate speech with target-domain emotion and speaker identity.

\subsection{Layer Selection for Emotion and Speaker Adaptation}
\label{layer_select}
Based on the weighted-sum learning for speaker identification and emotion recognition, $\bm{W}_e$ and $\bm{W}_s$ reflect the contribution of different Transformer layers in the TTS model to controlling emotion and speaker identity in the generated speech. Since we aim to learn the control of both emotion and speaker attributes simultaneously, we first compute the mean of $\bm{W}_e$ and $\bm{W}_s$, denoted as $\bm{W}_m$:
\begin{equation}
\label{mean}
    \bm{W}_m=\frac{\bm{W}_e+\bm{W}_s}{2}.
\end{equation}

We then select the two layers corresponding to the highest and lowest mean values for fine-tuning on the target-domain data. The layer with the lowest weight, which contains minimal emotion and speaker-related knowledge, is fine-tuned to strengthen its contribution to controlling the speaker and emotion of the generated speech. In contrast, the layer with the highest weight, which captures the most emotion and speaker information, is leveraged to maximize its utility. By freezing the remaining layers, we preserve the pre-trained knowledge, thereby minimizing the risk of catastrophic forgetting, which could otherwise undermine the TTS model's performance, especially in terms of word accuracy. This strategy strikes a balance between enabling emotion and speaker adaptation in the characteristic-specific layers and maintaining the model's strong foundational knowledge, leading to an effective and efficient adaptation for generating speech with the desired target-domain emotion and speaker expression.

\subsection{Characteristic-Specific Partial Fine-Tuning}
Section \ref{layer_select} selects two Transformer layers for emotion and speaker adaptation based on the characteristic-specific weights derived in Section \ref{superb_ft}. Since the utterance-level emotion and speaker classification explored in Section \ref{superb_ft} is theoretically independent of the speech content \cite{expressive_vc_first}, the speaker and emotion controlling capabilities of the characteristic-specific selected layers are also cross-lingual and content-agnostic. When working with target-domain datasets containing specific emotional expressions and speaker characteristics, we only need to freeze the remaining parameters and fine-tune the selected two Transformer layers on the target-domain TTS dataset, as shown in Figure ~\ref{proposed} (b). This allows the model to efficiently learn to generate speech with the emotional expression and speaker identity of the target domain. The cross-lingual and content-agnostic property of our approach allows us to complete the preparatory work in Sections \ref{superb_ft} and \ref{layer_select} using open-source data with speaker and emotion annotations. We can then directly apply the selected modules to target-domain TTS fine-tuning, eliminating the need to repeat the exploration of Sections \ref{superb_ft} and \ref{layer_select} on the target-domain dataset.

\section{Experiment Setup}
To validate our approach, we conduct target-domain emotion and speaker adaptation experiments using four open-source pre-trained codec language TTS models.

\subsection{Dataset}
\label{dataset}
We collect 11 public English emotional speech datasets: CREMA-D \cite{cao2014crema}, EMNS \cite{EMNS_corpus}, EmoV-DB \cite{emov}, eNTERFACE \cite{enterface}, IEMOCAP \cite{busso2008iemocap}, JL-Corpus \cite{jlcorpus}, MEAD \cite{kaisiyuan2020mead}, MELD \cite{meld}, RAVDESS \cite{ravdess}, TESS \cite{tess}, and MSP \cite{martinez2020msp}. 
To ensure consistency and class balance, we retain only emotions with over 8,000 samples, resulting in eight emotion labels: angry, happy, sad, fear, disgust, neutral, surprise, and contempt. The final combined dataset comprises 2,060 speakers and 244 hours of speech data, partitioned into training and testing sets in a $9:1$ ratio. For valid speaker adaptation evaluation, we ensure that all test-set speakers and emotions are also present in the training set.
Furthermore, alongside the English corpora, we incorporate Chinese emotional speech data from ESD \cite{esd} to evaluate the effectiveness and robustness of the identified characteristic-specific layer weights in cross-lingual and cross-dataset domain adaptation tasks.

\subsection{Model and Training Setup}
As outlined in our methodology, we first utilize a weighted-sum approach to analyze the characteristic-specific layer weights for emotion and speaker control. Based on these weights, we explicitly select and fine-tune specific Transformer layers for target-domain adaptation. This section details the selected pre-trained TTS models and their configurations.

\subsubsection{Pre-trained Codec Language TTS Model Setup}
We select four representative open-source codec language TTS models: GPT-SoVITS\footnote{\url{https://github.com/RVC-Boss/GPT-SoVITS}}, VALLE-X\footnote{\url{https://github.com/X-LANCE/SLAM-LLM/tree/main/examples/vallex}}, CosyVoice\footnote{\url{https://github.com/FunAudioLLM/CosyVoice}}, and Fun-CosyVoice3.0\footnote{\url{https://huggingface.co/FunAudioLLM/Fun-CosyVoice3-0.5B-2512}}. Crucially, these models comprehensively cover the primary speech information modeling paradigms in current LLM-based TTS: Self-supervised learning (SSL) tokens, full-information acoustic tokens, speech tokens with extra acoustic conditions, and ASR tokens, respectively.

\textbf{GPT-SoVITS}: This model employs RoBERTa \cite{robeta} for text encoding and HuBERT \cite{hsu2021hubert} with vector quantization to extract discrete SSL speech tokens. The concatenated text and speech representations are processed by a 24-layer Transformer, followed by a VITS decoder \cite{kim2021conditional} to reconstruct the speech.

\textbf{VALLE-X}: This model utilizes Encodec \cite{encodec} to extract full-information acoustic tokens. Both speech and text tokens are embedded and input into a 24-layer Transformer. A non-autoregressive module and the Encodec decoder are then used for speech reconstruction.

\textbf{CosyVoice}: Text is encoded using a Conformer \cite{gulati2020conformer}, while speech is quantized into discrete tokens via SenseVoice \cite{an2024funaudiollm}, incorporating extra acoustic conditions. The concatenated embeddings pass through a 14-layer Transformer, and the output is synthesized back into speech using Flow Matching and a HiFi-GAN vocoder \cite{mehta2024matcha}.

\textbf{Fun-CosyVoice3.0}: This model leverages a multi-task supervised speech tokenizer to convert speech into discrete ASR tokens. For our experiments, we use the 0.5B-parameter variant, which inputs the concatenated speech and text representations into a 24-layer Transformer, similarly employing Flow Matching and HiFi-GAN for final speech generation.

\subsubsection{Characteristic-Specific Analysis Setup}
\label{ss_ftm}
As outlined in Section~\ref{downstream_models}, we utilize lightweight downstream modules to perform emotion recognition and speaker identification, ensuring that the training focus remains effectively on learning the layer weights. Each downstream module consists of three convolutional layers with a kernel size of 5, a pooling kernel size of 5, and an internal dimension of 256. 
During this characteristic-specific analysis, the four pre-trained codec language TTS models are entirely frozen. We optimize only the layer weights and the downstream modules in a multi-task learning manner. This training process is conducted on the 244-hour English emotional dataset for 75 epochs using a single NVIDIA 3090 GPU. We employ the Adam optimizer \cite{kingma2014adam}, with the learning rate linearly warming up from 0 to 5e-4 over the first 8\% of the training steps, followed by a gradual decay to 0.

\subsubsection{Target-Domain Adaptation Setup}
To ensure a fair comparison, all evaluated fine-tuning strategies (i.e., partial, full, and other parameter-efficient approaches) follow consistent configurations for each respective codec language TTS model. The four pre-trained models are fine-tuned on the target domain for 10 epochs. To maintain training stability, the peak learning rates are set to approximately 5\% of their official pre-training values: 5e-4 for GPT-SoVITS, 2.5e-5 for VALLE-X, 1e-4 for CosyVoice, and 1e-5 for Fun-CosyVoice3.0. We utilize the Adam optimizer, with the learning rate linearly warming up from 0 to its peak over the first 8\% of the training steps before gradually decaying to 0. This configuration effectively facilitates target-domain adaptation while mitigating the loss of pre-trained foundational knowledge.

\subsection{Baseline Methods}
We compare our proposed \our{} method against two standard adaptation approaches: Low-Rank Adaptation (LoRA) \cite{hu2022lora} and full fine-tuning. 
LoRA is a prominent parameter-efficient fine-tuning (PEFT) technique widely adopted for large language and TTS models \cite{lou2024stylespeech}. It freezes the pre-trained weights and introduces lightweight trainable matrices to adapt to new domains. For a fair evaluation, we precisely adjust the rank of LoRA so that its total number of trainable parameters aligns with that of our partial fine-tuning strategy. 
Conversely, full fine-tuning updates all parameters across the entire model to maximize the learning of domain-specific knowledge. To guarantee strict comparability, we maintain consistent hyperparameters across all methods, including batch sizes, training epochs, and the learning rates detailed in the previous section.

\subsection{Evaluation}
\label{evaluation_setup}
For the initial characteristic-specific analysis, we use classification accuracy to evaluate the sentence-level emotion recognition and speaker identification tasks. 
To assess the target-domain adaptation performance, we evaluate the fine-tuned TTS models using unseen text from the test set. Specifically, we input the target test text alongside a reference speech prompt from the training set that exhibits the desired emotional expression and speaker identity. The generated speech is then compared against the ground-truth test-set audio to evaluate both characteristic similarity and content preservation.

\textbf{Objective Evaluation:} We evaluate Speaker Similarity (SS), Emotion Representation Similarity (ERS), Word Error Rate (WER) for English, and Character Error Rate (CER) for Chinese. SS is computed as the cosine similarity between speaker embeddings extracted from the generated and ground-truth speech using Resemblyzer\footnote{\url{https://github.com/resemble-ai/Resemblyzer}}. Similarly, ERS is measured via the cosine similarity of representations extracted by Emotion2vec+large\footnote{\url{https://github.com/ddlBoJack/emotion2vec}}. To evaluate intelligibility and monitor potential catastrophic forgetting, we calculate the English WER using Whisper$_{\text{Large V3}}$\footnote{\url{https://github.com/openai/whisper}} and the Chinese CER using Paraformer-zh\footnote{\url{https://github.com/modelscope/FunASR}}.

\textbf{Subjective Evaluation:} We conduct subjective listening tests to evaluate Emotion Similarity (EMOS), Speaker Similarity (SMOS), and overall Naturalness. Fifteen graduate students with expertise in speech processing participated in the evaluation. Scores were assigned on a Mean Opinion Score (MOS) scale from 1 to 5 with 0.5-point intervals.

\section{Experiment Results}
To comprehensively evaluate our approach, we first validate the four pre-trained codec language TTS models as encoders for emotion recognition and speaker identification. This establishes the reliability of the derived layer-wise characteristic-specific weights. Next, we compare \our{} against baseline methods to demonstrate its efficacy in emotion and speaker adaptation. We further assess cross-lingual and cross-dataset robustness on Chinese corpora, and evaluate adaptation efficiency through training speed comparisons. Finally, we conduct ablation studies to determine the optimal selection and number of fine-tuned layers.

\subsection{Performance on Speaker and Emotion Recognition}
\label{superb_exp}
To validate the effectiveness of the layer-wise weights identified in the characteristic-specific analysis, we compare the performance of Whisper \cite{radford2023robust}, HuBERT \cite{hsu2021hubert}, WavLM \cite{chen2022wavlm}, and the four codec language TTS models on weighted-sum speaker and emotion recognition tasks (using the 244-hour English emotional dataset), as presented in Table~\ref{s_e_finetune}. Note that Whisper utilizes only its encoder for this evaluation.

\begin{table}[htp]
\setlength{\tabcolsep}{3pt}
\setlength{\extrarowheight}{1.pt}
\centering
\small
\caption{Accuracy of different pre-trained models on emotion recognition and speaker identification.}
\begin{tabular}{lccc}
    \toprule
    \multirow{2}{*}{\shortstack{Pre-trained\\Model}} & \multirow{2}{*}{\shortstack{Param.\\(M)}} & \multirow{2}{*}{\shortstack{Speaker\\Accuracy (\%)} $\uparrow$}  & \multirow{2}{*}{\shortstack{Emotion\\Accuracy (\%)} $\uparrow$}  \\
    \\
    \cline{1-4}
    Fbank & 0 & 84.33 & 43.64   \\
    \cline{1-4}
    Whisper$_\text{small}$ & 86.89 & 88.45 & 67.13   \\
    HuBERT$_\text{base}$ & 94.68 & 89.14 & 67.99  \\
    WavLM$_\text{base}$  & 94.68 & 90.83 & 69.01   \\
    Whisper$_\text{medium}$  & 305.38 & 92.75 & 69.84   \\
    HuBERT$_\text{large}$  & 316.61 & 93.25 & 71.66 \\
    WavLM$_\text{large}$  & 316.61 & 93.59 & \textbf{72.61}   \\
    \cline{1-4}
    GPT-SoVITS & 77.46 & 89.02 & 63.24  \\
    VALLE-X & 362.09 & 93.91 & 66.03  \\
    Fun-CosyVoice3.0 & 506.15 & 86.87 & 72.44  \\
    CosyVoice & 310.72 &\textbf{94.97} & 70.48  \\
    \bottomrule
\end{tabular}
\label{s_e_finetune}
\end{table}

\begin{table*}[htp]
\centering
\small
\caption{Comparison of fine-tuning strategies for adapting pre-trained TTS models to generate speech with target-domain speaker and emotion expressions. Param. indicates the trainable/total parameters. \textbf{Bold} represents the best score, and \underline{underline} denotes the second-best score.}
\resizebox{\linewidth}{!}{
\setlength\tabcolsep{3pt}
\setlength{\extrarowheight}{1.pt}
\begin{tabular}{l|c|ccc|c|ccc|c|ccc|c|ccc}
    \hline
    \multirow{2.5}{*}{Method} & \multicolumn{4}{c|}{GPT-SoVITS} & \multicolumn{4}{c|}{VALLE-X} & \multicolumn{4}{c|}{CosyVoice} & \multicolumn{4}{c}{Fun-CosyVoice3.0} \\
    \cline{2-17}
    &Param.&\makecell{SS\\(\%)$\uparrow$}&\makecell{ERS\\(\%)$\uparrow$}&\makecell{WER\\(\%)$\downarrow$}&Param.&\makecell{SS\\(\%)$\uparrow$}&\makecell{ERS\\(\%)$\uparrow$}&\makecell{WER\\(\%)$\downarrow$}&Param.&\makecell{SS\\(\%)$\uparrow$}&\makecell{ERS\\(\%)$\uparrow$}&\makecell{WER\\(\%)$\downarrow$}&Param.&\makecell{SS\\(\%)$\uparrow$}&\makecell{ERS\\(\%)$\uparrow$}&\makecell{WER\\(\%)$\downarrow$}\\
    \hline
    Origin & 0/77.5 & 64.0 & 77.7 & \textbf{8.0} & 0/362.1 & 84.2 & 85.6 & 8.4 & 0/310.7 & 88.9 & 89.8 & \textbf{8.4} & 0/506.2 & 91.2 & 92.0 & 4.0 \\
    \hline
    Full FT & 77.5/77.5 & 69.4 & \textbf{83.9} & 10.3 & 362.1/362.1 & \textbf{90.2} & 93.0 & 9.5 & 310.7/310.7 & 92.5 & \underline{95.6} & 25.1 & 506.2/506.2 & 94.5 & \textbf{97.0} & 12.1 \\
    \hline
    \multirow{3}{*}{LoRA FT} & 3.1/80.6 & 68.7 & 82.4 & 14.7 & 12.7/374.8 & 89.2 & 91.7 & 9.5 & 13.8/324.5 & 91.8 & 93.3 & 20.6 & 14.5/520.7 & 93.0 & 94.1 & 9.2 \\
    & 6.2/83.7 & 68.7 & 82.2 & 12.1 & 25.1/387.2 & 89.5 & 91.9 & 9.9 & 27.5/338.3 & 92.0 & 93.1 & 18.1 & 29.0/535.2 & 93.1 & 94.0 & 8.5 \\
    & 9.3/86.8 & 68.7 & 82.4 & 11.8 & 37.5/399.6 & 89.5 & 92.2 & 9.6 & 41.3/352.0 & 91.9 & 93.6 & 19.3 & 43.5/549.7 & 93.3 & 94.3 & 8.9 \\
    \hline
    First Half FT & 37.8/77.5 & \underline{69.4} & 83.2 & 13.7 & 151.2/362.1 & 89.6 & 92.9 & 8.4 & 95.5/310.7 & 92.3 & 94.2 & 22.6 & 178.9/506.2 & 93.5 & 95.2 & 11.5 \\
    Second Half FT & 37.8/77.5 & 69.4 & 82.8 & 14.5 & 151.2/362.1 & 89.4 & 92.3 & 9.8 & 95.5/310.7 & 92.4 & \textbf{95.6} & 15.7 & 178.9/506.2 & 93.7 & 96.6 & 7.5 \\
    Shallowest Two & 6.3/77.5 & 68.9 & 82.5 & 10.4 & 25.2/362.1 & 89.0 & 92.1 & 8.4 & 27.3/310.7 & 92.3 & 94.7 & 14.7 & 29.8/506.2 & 93.6 & 95.5 & 6.8 \\
    Deepest Two & 6.3/77.5 & 68.5 & 82.4 & 13.7 & 25.2/362.1 & 89.4 & 92.5 & 9.8 & 27.3/310.7 & 92.3 & 94.0 & 15.7 & 29.8/506.2 & 93.5 & 95.0 & 7.2 \\
    \hline
    Lowest Two FT & 6.3/77.5 & 68.3 & 82.2 & 9.5 & 25.2/362.1 & 89.7 & 93.0 & \underline{8.0} & 27.3/310.7 & 92.3 & 94.8 & 12.3 & 29.8/506.2 & 93.6 & 95.8 & \underline{4.2} \\
    Highest Two FT & 6.3/77.5 & 69.1 & 83.0 & 12.4 & 25.2/362.1 & 89.7 & \underline{93.1} & 10.0 & 27.3/310.7 & \underline{92.5} & 95.3 & 14.6 & 29.8/506.2 & \underline{94.6} & 96.5 & 6.5 \\
    \textbf{\our{}} (Ours) & 6.3/77.5 & \textbf{69.9} & \underline{83.2} & \underline{8.5} & 25.2/362.1 & \underline{90.2} & \textbf{94.2} & \textbf{7.0} & 27.3/310.7 & \textbf{92.6} & 95.5 & \underline{9.2} & 29.8/506.2 & \textbf{94.8} & \underline{96.8} & \textbf{3.8} \\
    \hline
\end{tabular}
}
\label{tab:results}
\end{table*}

The results indicate that generative codec TTS models exhibit capabilities comparable to SOTA BERT-style pre-trained models in perception tasks. However, performance varies significantly based on tokenization strategies. GPT-SoVITS achieves competitive speaker identification but yields moderate English emotion recognition accuracy, primarily because it relies on a Chinese-pretrained cn-HuBERT for token extraction. VALLE-X extracts full-information acoustic tokens from the first layer of Encodec \cite{borsos2023audiolm}, capturing rich acoustic information that significantly boosts speaker accuracy. However, this shallow acoustic representation limits its effectiveness on semantic-heavy emotion tasks \cite{viola}.
Conversely, Fun-CosyVoice3.0 exhibits a stark contrast in its performance. Utilizing ASR tokens, it achieves the highest emotion recognition accuracy (72.44\%) among all evaluated TTS models and closely rivals the SOTA WavLM$_\text{large}$ (72.61\%). This confirms that ASR tokens effectively retain crucial semantic information. However, this reliance on ASR tokens is also the primary reason for its poor speaker identification performance. ASR tokens are inherently designed to discard speaker-specific acoustic details in favor of linguistic content, resulting in a significant drop in speaker accuracy (86.87\%), which is the lowest among the evaluated pre-trained models. Finally, CosyVoice achieves the strongest balanced performance across both domains. Although it similarly employs ASR tokens to preserve semantics for competitive emotion recognition (70.48\%), it overcomes the inherent loss of acoustic information by explicitly incorporating an extra speaker embedding as input to its language model. This architectural choice allows CosyVoice to maintain the highest speaker accuracy (94.97\%).
While direct comparisons are bounded by different pre-training datasets and model architectures, these results confirm the reliability of our layer-wise weight analysis and highlight the broader potential of autoregressive generative models in universal speech perception tasks.

\subsection{Comparison with Reference Methods}
From the experiments in Section \ref{superb_exp}, we observe that the characteristic-specific weights identified through weighted-sum analysis provide a valuable reference for selecting parameters for emotion and speaker adaptation of codec language TTS models. In this section, we compare our proposed method with eight fine-tuning strategies across the four models: full fine-tuning, LoRA fine-tuning (with trainable parameters matched to our partial fine-tuning), fine-tuning the first half of the Transformer, fine-tuning the second half of the Transformer, fine-tuning the shallowest two layers of the Transformer, fine-tuning the deepest two layers of the Transformer, fine-tuning the two layers with the highest characteristic-specific weights, and fine-tuning the two layers with the lowest characteristic-specific weights. The experimental results are presented in Table~\ref{tab:results}.

\subsubsection{Comparison with Full Fine-Tuning}
The results in Table~\ref{tab:results} indicate that full fine-tuning effectively adapts all four TTS models to the target data domain, resulting in significant improvements in SS and ERS scores. However, it also leads to noticeable catastrophic knowledge forgetting, as evidenced by the sharp increase in WER. This phenomenon is particularly evident in the large-scale Fun-CosyVoice3.0 model: while full fine-tuning pushes its ERS to the absolute highest (97.0\%), its WER severely degrades from 4.0\% to 12.1\%. In contrast, our proposed method achieves SS and ERS scores comparable to or even surpassing those of full fine-tuning while using a substantially smaller number of trainable parameters and minimizing knowledge forgetting (e.g., maintaining a low WER of 3.8\% for Fun-CosyVoice3.0). This highlights the importance of analyzing the characteristics of each module within the model to determine its relevance for target adaptation. Such analysis is crucial for identifying which modules to fine-tune, enabling effective target-domain adaptation with reduced training costs, as demonstrated by the findings in \cite{intro_cnn}.

\subsubsection{Comparison with LoRA Fine-Tuning}
We configure the parameters of the inserted LoRA modules (applied across all Transformer layers) to match the parameter counts for fair comparison. The results in Table~\ref{tab:results} show that as the LoRA parameters increase, the adapted TTS models exhibit improved speaker cloning and emotional expression capabilities, but this improvement often comes at the cost of a higher WER. This occurs because LoRA indirectly updates the model via matrix decomposition, leaving the original parameters unchanged while aligning the model's intermediate feature space with the target dataset. However, this alignment can lead to challenges such as catastrophic forgetting or overfitting. For instance, even with scaled-up LoRA ranks on Fun-CosyVoice3.0, the performance ceiling (e.g., an SS of 93.3\%) remains lower than our targeted layer selection strategy (SS 94.8\%). In contrast, our method achieves superior performance without introducing additional architecture modules during fine-tuning, demonstrating that fully freezing pre-trained parameters and blindly adding trainable matrices is not the optimal strategy. Instead, it is crucial to carefully select which inherent parameters to fine-tune based on the specific demands of the target task or data adaptation.

\subsubsection{Comparison with other Partial Fine-Tuning}
To verify the effectiveness of characteristic-specific weights in guiding partial fine-tuning, we compared the performance of various partial fine-tuning strategies, as shown in Table~\ref{tab:results}. We began by fine-tuning different parts of the Transformer, including the first half, the second half, the two shallowest layers, and the two deepest layers. The results reveal that simply increasing the number of trainable parameters does not necessarily improve performance and may even reduce the effectiveness of fine-tuning. A stark contrast is observed in Fun-CosyVoice3.0: fine-tuning half of the network updates nearly 179M parameters but yields a WER of at least 7.5\%, whereas our method updates only roughly 30M parameters yet achieves better speaker similarity (94.8\%) and significantly lower knowledge forgetting (WER 3.8\%). Furthermore, we compared fine-tuning the two layers with the highest weights and the two layers with the lowest weights. Our proposed approach, which combines fine-tuning the layers with the highest and lowest weights, proved to be more effective. Specifically, our \our{} jointly fine-tunes two carefully selected layers: one with the best adaptation performance but moderate catastrophic forgetting (highest), and the other with minimal catastrophic forgetting but moderate adaptation performance (lowest). By complementing each other, fine-tuning these two layers enables the model to achieve excellent adaptation capability with minimal knowledge forgetting, resulting in outstanding overall performance.

\begin{table*}[htp]
\setlength\tabcolsep{2pt}
\setlength{\extrarowheight}{1.5pt}
\centering
\small
\caption{Subjective results evaluated by 15 listeners, with 95\% confidence intervals computed from the t-test. SMOS, EMOS, and NMOS evaluate speaker similarity, emotion similarity, and speech naturalness, respectively. \textbf{Bold} represents the best score.}
\vspace{-0.2cm}
\resizebox{\linewidth}{!}{
\begin{tabular}{l|ccc|ccc|ccc|ccc}
    \hline
    \multirow{2}{*}{Method} & \multicolumn{3}{c|}{GPT-SoVITS} & \multicolumn{3}{c|}{VALLE-X} & \multicolumn{3}{c|}{CosyVoice} & \multicolumn{3}{c}{Fun-CosyVoice3.0} \\
    \cline{2-13}
    &  \makecell[c]{SMOS$\uparrow$}& \makecell{EMOS$\uparrow$} & \makecell{NMOS$\uparrow$} &  \makecell[c]{SMOS$\uparrow$} & \makecell{EMOS$\uparrow$} & \makecell{NMOS$\uparrow$} &  \makecell[c]{SMOS$\uparrow$} & \makecell{EMOS$\uparrow$} & \makecell{NMOS$\uparrow$} &  \makecell[c]{SMOS$\uparrow$} & \makecell{EMOS$\uparrow$} & \makecell{NMOS$\uparrow$} \\
    \hline
    Origin & 3.12$\pm$0.08 & 3.35$\pm$0.09 & \textbf{4.15$\pm$0.06} & 3.65$\pm$0.08 & 3.52$\pm$0.09 & 4.12$\pm$0.07 & 3.80$\pm$0.09 & 3.75$\pm$0.08 & \textbf{4.25$\pm$0.06} & 3.95$\pm$0.08 & 3.88$\pm$0.09 & 4.35$\pm$0.05 \\
    Full FT & 3.55$\pm$0.09 & \textbf{3.82$\pm$0.11} & 3.52$\pm$0.12 & 3.92$\pm$0.09 & 4.05$\pm$0.10 & 3.85$\pm$0.11 & 4.15$\pm$0.08 & \textbf{4.30$\pm$0.09} & 2.85$\pm$0.14 & 4.28$\pm$0.07 & \textbf{4.45$\pm$0.08} & 3.45$\pm$0.12 \\
    LoRA FT & 3.48$\pm$0.10 & 3.75$\pm$0.08 & 3.25$\pm$0.11 & 3.85$\pm$0.08 & 3.95$\pm$0.09 & 3.80$\pm$0.08 & 4.08$\pm$0.10 & 4.15$\pm$0.09 & 3.12$\pm$0.12 & 4.15$\pm$0.09 & 4.25$\pm$0.08 & 3.95$\pm$0.09 \\
    \textbf{\our{}} & \textbf{3.61$\pm$0.07} & 3.78$\pm$0.08 & 4.02$\pm$0.08 & \textbf{3.95$\pm$0.07} & \textbf{4.12$\pm$0.08} & \textbf{4.20$\pm$0.06} & \textbf{4.18$\pm$0.07} & 4.28$\pm$0.08 & 4.15$\pm$0.08 & \textbf{4.35$\pm$0.06} & 4.42$\pm$0.07 & \textbf{4.40$\pm$0.06} \\
    \hline
\end{tabular}
}
\vspace{-0.3cm}
\label{mos_table}
\end{table*}

\subsubsection{Subjective Evaluation}
To validate whether the objective metrics align with human perception, we conducted a Mean Opinion Score (MOS) listening test. We engaged 15 graduate students as listeners to evaluate the generated speech samples across three dimensions: Speaker MOS (SMOS) for voice cloning similarity, Emotion MOS (EMOS) for emotional similarity, and Naturalness MOS (NMOS) for overall speech quality and intelligibility. The subjective results, along with their 95\% confidence intervals computed from the t-test, are summarized in Table~\ref{mos_table}.

The subjective results perfectly corroborate our objective findings. The Original models possess high naturalness (NMOS) but lack adaptation capability, resulting in the lowest SMOS and EMOS. While Full FT significantly enhances speaker and emotion similarity, it suffers a severe degradation in NMOS. This is particularly noticeable in CosyVoice, where the NMOS drops drastically to 2.85, completely aligning with the catastrophic knowledge forgetting implied by its 25.1\% WER. Similarly, LoRA FT struggles to maintain a balance, often compromising speech intelligibility to fit the target domain. In stark contrast, our proposed \our{} method consistently achieves the best balance across all models. It matches or exceeds the SMOS and EMOS of full fine-tuning while tightly preserving speech naturalness. Remarkably, on the massive Fun-CosyVoice3.0 model, our method achieves an NMOS of 4.40, effectively surpassing the original model's 4.35. This proves that our layer-wise characteristic-specific selection not only prevents catastrophic forgetting objectively but also delivers highly natural, expressive, and accurate speech from a human auditory perspective.

\subsection{Transferability of Characteristic-Specific Weights Across Datasets}
In practice, obtaining emotion and speaker annotations for the target domain dataset is challenging. This limitation highlights the need for our emotion and speaker characteristic-specific weights to be transferable across datasets. To verify this, we fine-tune TTS models to the Chinese emotional dataset using the layer-wise characteristic-specific weights learned from the English emotional dataset, as described in Section~\ref{dataset}. The results, presented in Table~\ref{cross_domain}, include LoRA fine-tuning configured with a parameter count equivalent to the 2-layer fine-tuning.
The results show that while English-pretrained models generally experience a slight performance drop when adapting to Chinese, Fun-CosyVoice3.0 exhibits exceptional cross-lingual robustness and achieves the lowest initial Character Error Rate (CER) among all models. After fine-tuning with our method, both SS and ERS improve significantly across all baselines. For GPT-SoVITS, the adaptation improves speaker and emotion metrics with only a minimal increase in CER. More importantly, for VALLE-X, CosyVoice, and the large-scale Fun-CosyVoice3.0, our method consistently achieves the highest SS and ERS compared to full fine-tuning and LoRA fine-tuning. Furthermore, our approach tightly preserves the CER relative to the original zero-shot models, effectively avoiding the noticeable degradation caused by full and LoRA fine-tuning. This is particularly evident in Fun-CosyVoice3.0, where the CER is successfully maintained at an impressive 1.2\%. These results confirm that the proposed layer-wise characteristic-specific weights are highly transferable and effective for cross-dataset adaptation in emotion and speaker control. Since the relationship between controlling utterance-level characteristics (such as emotional state and speaker identity) and the linguistic content is minimal, the characteristic-specific analysis only needs to be performed once on an open-source annotated dataset. This analysis can then be reliably applied to other datasets, enabling the model to adapt and generate speech with the target emotion and speaker characteristics without heavy retraining.

\begin{table}[htp]
\setlength\tabcolsep{1pt}
\setlength{\extrarowheight}{2.pt}
\centering
\small
\caption{Cross-dataset evaluation of characteristic-specific weight for emotion and speaker adaptation.}
\vspace{-0.2cm}
\resizebox{\linewidth}{!}{
\begin{tabular}{l|ccc|ccc|ccc|ccc}
    \hline
    \multirow{2}{*}{Method} & \multicolumn{3}{c|}{GPT-SoVITS} & \multicolumn{3}{c|}{VALLE-X} & \multicolumn{3}{c|}{CosyVoice} & \multicolumn{3}{c}{Fun-CosyVoice3.0} \\
    \cline{2-13}
    &  \makecell[c]{SS\\(\%)$\uparrow$}& \makecell{ERS\\(\%)$\uparrow$} & \makecell{CER\\(\%)$\downarrow$} &  \makecell[c]{SS\\(\%)$\uparrow$} & \makecell{ERS\\(\%)$\uparrow$} & \makecell{CER\\(\%)$\downarrow$} &  \makecell[c]{SS\\(\%)$\uparrow$} & \makecell{ERS\\(\%)$\uparrow$} & \makecell{CER\\(\%)$\downarrow$} &  \makecell[c]{SS\\(\%)$\uparrow$} & \makecell{ERS\\(\%)$\uparrow$} & \makecell{CER\\(\%)$\downarrow$} \\
    \hline
    Origin & 73.8 & 70.3 & \textbf{2.2} & 75.4 & 82.3 & \textbf{5.2} & 79.8 & 88.6 & \textbf{2.5} & 83.5 & 90.5 & \textbf{1.2} \\
    Full & \textbf{75.4} & \underline{80.9} & 3.3 & \underline{77.7} & \underline{86.1} & 7.6 & \underline{81.0} & \underline{91.2} & 3.0 & \underline{85.1} & \underline{93.9} & 3.9 \\
    LoRA & 74.8 & 78.5 & 3.4 & 76.9 & 84.2 & 7.4 & 80.2 & 89.9 & 2.7 & 84.7 & 92.8 & 2.1 \\
    \textbf{\our{}} & \underline{75.2} & \textbf{81.0} & \underline{2.5} & \textbf{78.0} & \textbf{87.0} & \underline{6.0} & \textbf{81.2} & \textbf{91.9} & \textbf{2.5} & \textbf{85.7} & \textbf{94.1} & \textbf{1.2} \\
    \hline
\end{tabular}
}
\label{cross_domain}
\end{table}

\subsection{Catastrophic Forgetting of Knowledge}
\begin{figure}[ht]
\centering
\includegraphics[width=8cm]{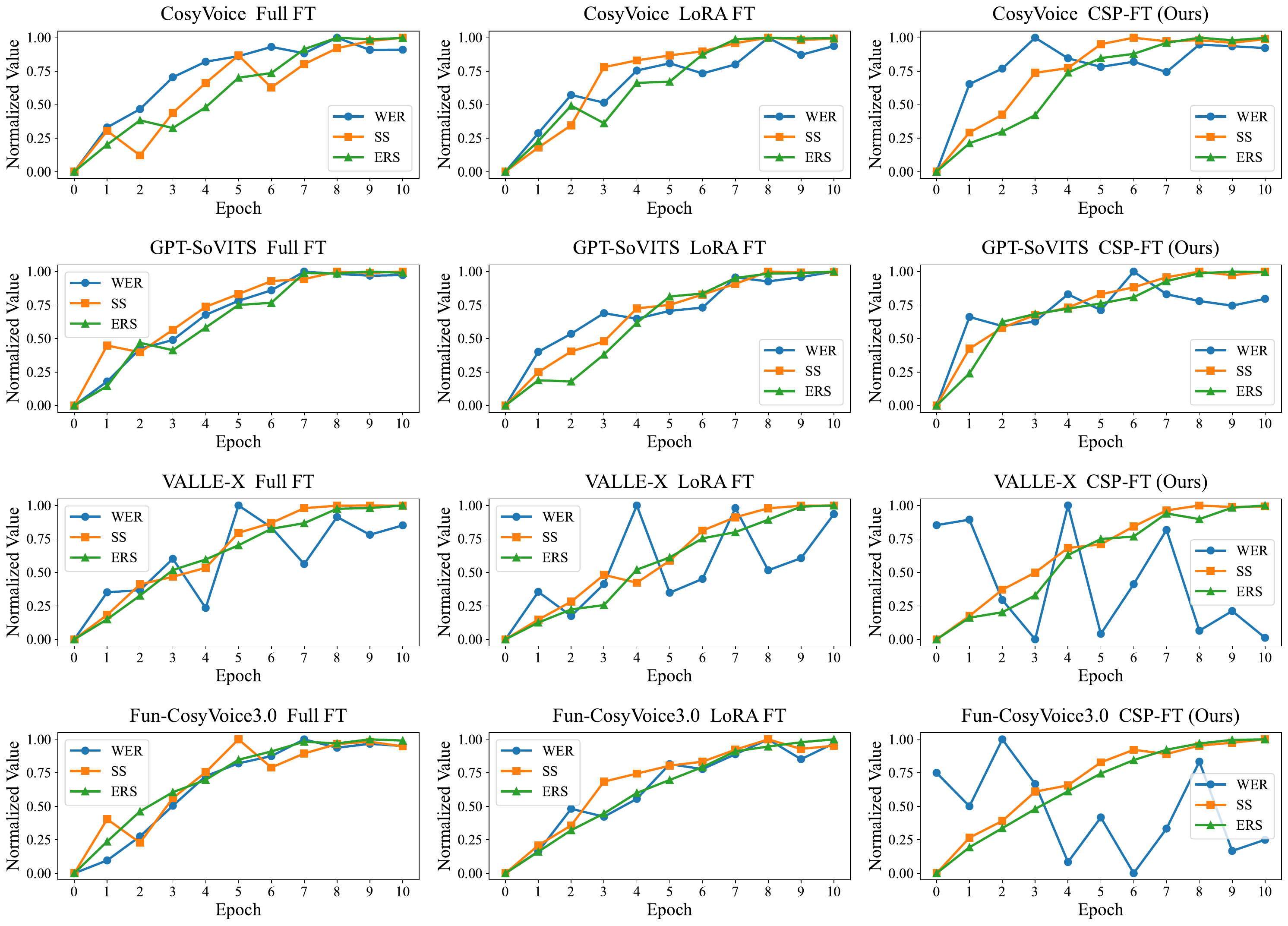}
\vspace{-0.5cm}
\caption{Performance trends of SS, ERS, and WER across fine-tuning epochs for different strategies.} 
\label{epoch}
\end{figure}

To further investigate the performance of different fine-tuning strategies in addressing the catastrophic knowledge forgetting problem, we evaluate performance across epochs during fine-tuning on the English emotional dataset. The results are shown in Figure~\ref{epoch}. To present various metrics within a single figure, we numerically normalize each metric $\bm{S}$ as:

\begin{equation} \bm{S}_\text{norm} = \frac{\bm{S} - \min(\bm{S})}{\max(\bm{S}) - \min(\bm{S})}. \end{equation}

From the first and second columns, which depict the results of full fine-tuning and LoRA fine-tuning for the four models, we observe that as the number of adaptation epochs increases, SS and ERS gradually rise and then plateau, while WER steadily increases. 
This occurs because the content knowledge in the target domain is much smaller than that in the pre-training domain, leading to profound knowledge forgetting.
In contrast, the last column shows that our method achieves improved SS and ERS while stabilizing WER significantly earlier. Notably, for both VALLE-X and Fun-CosyVoice3.0, the WER exhibits bounded fluctuations during the adaptation process before ultimately settling at better results than their original unfinetuned states. These findings demonstrate that our method enables the models to adapt effectively to generate speech with target emotion and speaker characteristics while actively minimizing the decline in word accuracy. In other words, it significantly alleviates the problem of catastrophic forgetting across various model architectures and parameter scales.

\begin{figure}[ht]
\centering
\includegraphics[width=\linewidth]{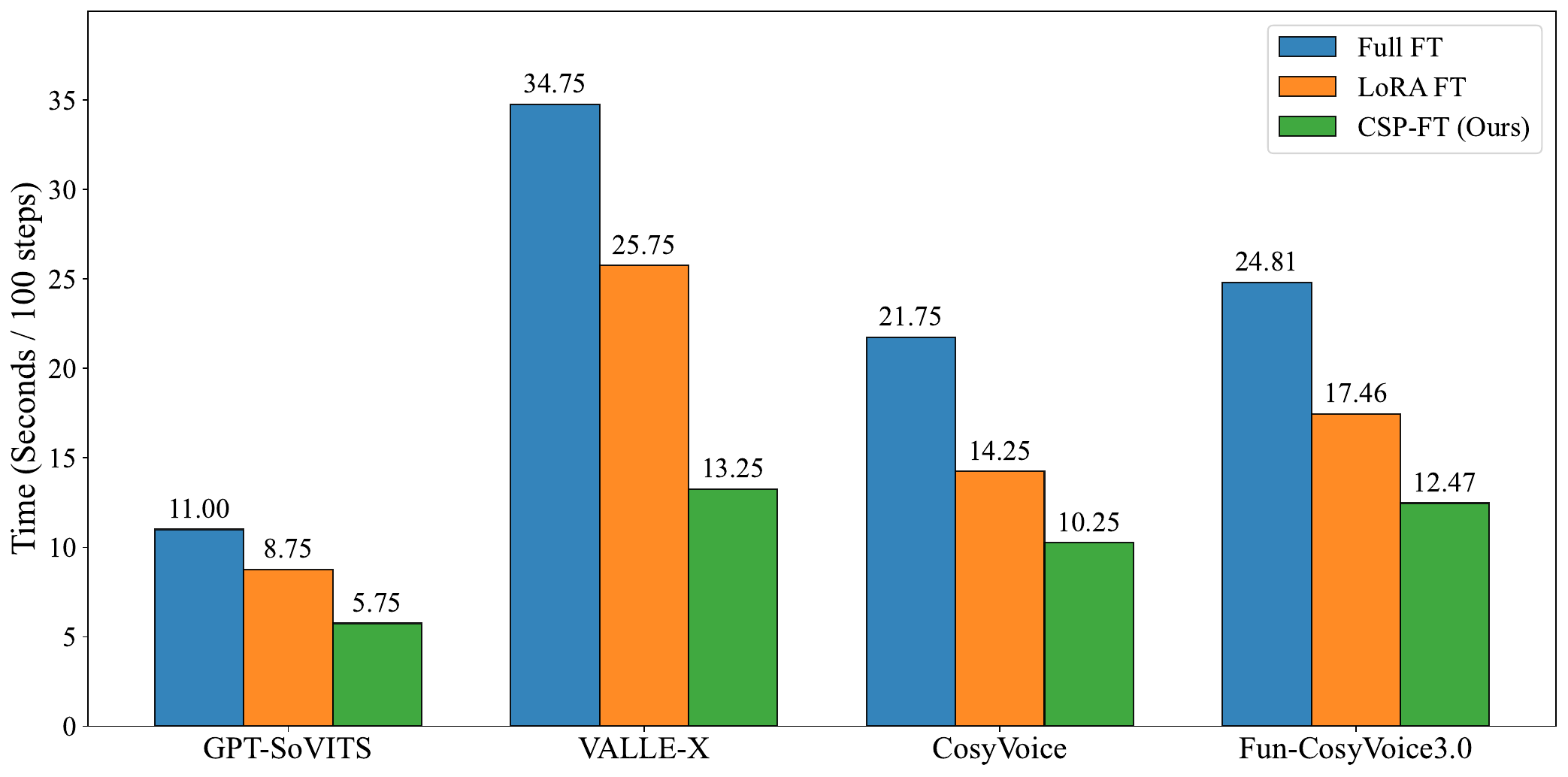}
\vspace{-0.4cm}
\caption{Time consumption of fine-tuning 100 steps using different fine-tuning methods.} 
\label{time_speed}
\end{figure}

\subsection{Fine-Tuning Speed}
Figure~\ref{time_speed} compares the online adaptation time for 100 optimization steps across full fine-tuning, parameter-matched LoRA, and our characteristic-specific partial fine-tuning. Because our highly transferable layer-wise analysis is performed once offline, this comparison strictly measures target-domain adaptation efficiency. Across all four models, our approach consistently outperforms both baselines. By selectively fine-tuning only the most relevant layers, our method significantly reduces the computational overhead of full backpropagation. Compared to full fine-tuning, it achieves speedups of 1.91$\times$ on GPT-SoVITS, 2.62$\times$ on VALLE-X, 2.12$\times$ on CosyVoice, and 1.99$\times$ on Fun-CosyVoice3.0. Crucially, as discussed in previous sections, this targeted parameter reduction not only accelerates training but also serves as a safeguard against the catastrophic forgetting and severe WER degradation seen in full fine-tuning. Furthermore, our strategy demonstrates a clear computational advantage over LoRA. While LoRA limits adaptation capability through indirect updates, it also introduces additional matrix multiplications into the computational graph. By directly updating inherent parameters without any supplementary architectural overhead, our method achieves faster training throughput while maintaining superior synthesis fidelity.

\begin{table}[ht]
\centering
\small
\setlength\tabcolsep{1.0pt}
\setlength{\extrarowheight}{2.5pt}
\caption{Performance comparison across different numbers of fine-tuned layers within our method.}
\vspace{-0.2cm}
\resizebox{\linewidth}{!}{
\begin{tabular}{l|ccc|ccc|ccc|ccc}
\hline
\multirow{2}{*}{Method} & \multicolumn{3}{c|}{GPT-SoVITS} & \multicolumn{3}{c|}{VALLE-X} & \multicolumn{3}{c|}{CosyVoice} & \multicolumn{3}{c}{Fun-CosyVoice3.0} \\
    \cline{2-13}
 &  \makecell[c]{SS\\(\%)$\uparrow$}& \makecell{ERS\\(\%)$\uparrow$} & \makecell{WER\\(\%)$\downarrow$} &  \makecell[c]{SS\\(\%)$\uparrow$} & \makecell{ERS\\(\%)$\uparrow$} & \makecell{WER\\(\%)$\downarrow$} &  \makecell[c]{SS\\(\%)$\uparrow$} & \makecell{ERS\\(\%)$\uparrow$} & \makecell{WER\\(\%)$\downarrow$} &  \makecell[c]{SS\\(\%)$\uparrow$} & \makecell{ERS\\(\%)$\uparrow$} & \makecell{WER\\(\%)$\downarrow$} \\
\hline
Origin & 64.0 & 77.7 & \textbf{8.0} & 84.2 & 85.6 & 8.4 & 88.9 & 89.8 & \textbf{8.4} & 91.2 & 92.0 & \underline{4.0} \\
\hline
Full FT & 69.4 & \textbf{83.9} & 10.3 & \textbf{90.2} & 93.0 & 9.5 & 92.5 & \underline{95.6} & 25.1 & 94.5 & \underline{97.0} & 12.1 \\
Ours+$\frac{5}{6}$ & 69.2 & 82.6 & 12.9 & 89.7 & 93.0 & 10.0 & 92.3 & 95.0 & 23.1 & 94.0 & 96.2 & 10.8 \\
Ours+$\frac{4}{6}$ & 68.9 & 82.3 & 13.1 & 89.0 & 92.5 & 9.1 & 91.5 & 94.5 & 25.7 & 93.5 & 95.5 & 9.5 \\
Ours+$\frac{3}{6}$ & 68.6 & 82.3 & 14.8 & 89.9 & \underline{94.0} & 9.0 & 91.7 & 94.4 & 22.7 & 93.8 & 95.8 & 8.0 \\
Ours+$\frac{2}{6}$ & \underline{69.6} & 83.2 & 12.9 & 89.7 & 93.5 & 8.6 & 92.3 & 94.8 & 17.8 & 94.2 & 96.3 & 6.5 \\
Ours+$\frac{1}{6}$ & 69.5 & 83.1 & 10.7 & 90.0 & 93.9 & \underline{8.0} & \underline{92.5} & \textbf{95.7} & 10.0 & \underline{94.6} & \textbf{97.0} & 5.2 \\
\our{} & \textbf{69.9} & \underline{83.2} & \underline{8.5} & \underline{90.2} & \textbf{94.2} & \textbf{7.0} & \textbf{92.6} & 95.5 & \underline{9.2} & \textbf{94.8} & 96.8 & \textbf{3.8} \\
\hline
\end{tabular}
}
\label{ablation}
\end{table}

\subsection{Ablation Study}
\subsubsection{Number of Selection}
The previous sections highlight the importance of layer-wise characteristic-specific weights in selecting partial fine-tuning modules for emotion and speaker adaptation. Building on this foundation, this section investigates the relationship between the number of layers selected based on our method and overall model performance. We gradually increase the number of partial fine-tuning layers according to the layer-wise characteristic-specific weights. Specifically, we start with 2 layers (those with the highest and lowest weights) and then incrementally increase the number of layers by $\frac{1}{6}$. For example, for an N-layer codec language TTS model, ``Ours+$\frac{1}{6}$" indicates selecting the $(1+\lfloor\frac{N}{12}\rfloor)$ highest and $(1+\lfloor\frac{N}{12}\rfloor)$ lowest weighted layers for fine-tuning, resulting in a total of $(2+\lfloor\frac{N}{6}\rfloor)$ layers, where $\lfloor\cdot\rfloor$ denotes the flooring function. The results are presented in Table~\ref{ablation}.

From Table~\ref{ablation}, we observe that as the number of fine-tuning layers increases from 2 to full fine-tuning, performance first decreases slightly and then improves. This trend reflects that while characteristic-specific weights are strongly correlated with final adaptation performance, this correlation weakens for layers with weights that fall in the middle range. Since our method begins by selecting layers with the highest and lowest task weights, configurations such as Ours+$\frac{3}{6}$ or Ours+$\frac{4}{6}$, which include more middle-weighted layers, may not fully align with the model's adaptation requirements. However, as the number of fine-tuned layers approaches full fine-tuning, performance shows a recovery. This U-shaped performance curve is remarkably pronounced in the large-scale Fun-CosyVoice3.0 model. While adding a few mid-weighted layers disrupts the optimal representation space, expanding towards full fine-tuning recovers the task metrics but severely exacerbates catastrophic forgetting, with the WER climbing continuously.

\subsubsection{Layer of Selection}
Quantitative analysis of the characteristic-specific weights reveals a distinct distribution across the network, with certain layers exhibiting scores close to the extreme values. To investigate the sensitivity of layer selection, we explore the effects of selecting the layers corresponding to the smallest, second smallest, or third smallest weights while keeping the layer with the highest weight fixed. Additionally, we test the effects of selecting the layers corresponding to the largest, second largest, or third largest weights, while fixing the layer with the smallest weight. The results are shown in Table~\ref{ablation2}.

The results show that modifying the selection of the layer with the small weights negatively impacts performance across almost all three metrics (with WER remaining flat for GPT-SoVITS). On the other hand, altering the selection of the layer with the large weights leads to a notable performance decline, especially in WER, on GPT-SoVITS, VALLE-X, and Fun-CosyVoice3.0. This happens because fine-tuning layers corresponding to the second or third largest weights fails to leverage the full learning potential of the best-performing layer, reducing fine-tuning effectiveness. For the 24-layer Fun-CosyVoice3.0, shifting away from the absolute highest-weighted layer results in a sharp drop in both adaptation capability and linguistic stability, reinforcing the necessity of precise layer localization in massive networks. Interestingly, for CosyVoice, selecting the second-largest weight layer marginally improves SS and ERS scores, despite a negligible WER variation (rounded to 9.2\% in both cases). We attribute this to the uneven distribution of emotional and speaker-related weights in CosyVoice, where the averaged weight provides a smoother result, enabling suboptimal but stable fine-tuning performance.

\begin{table}[ht]
\centering
\small
\setlength\tabcolsep{1.2pt}
\setlength{\extrarowheight}{2.5pt}
\caption{Ablation study on layer weight selection.}
\vspace{-0.2cm}
\resizebox{\linewidth}{!}{
\begin{tabular}{l|ccc|ccc|ccc|ccc}
\hline
\multirow{2}{*}{Method} & \multicolumn{3}{c|}{GPT-SoVITS} & \multicolumn{3}{c|}{VALLE-X} & \multicolumn{3}{c|}{CosyVoice} & \multicolumn{3}{c}{Fun-CosyVoice3.0} \\
    \cline{2-13}
 &  \makecell[c]{SS\\(\%)$\uparrow$}& \makecell{ERS\\(\%)$\uparrow$} & \makecell{WER\\(\%)$\downarrow$} &  \makecell[c]{SS\\(\%)$\uparrow$} & \makecell{ERS\\(\%)$\uparrow$} & \makecell{WER\\(\%)$\downarrow$} &  \makecell[c]{SS\\(\%)$\uparrow$} & \makecell{ERS\\(\%)$\uparrow$} & \makecell{WER\\(\%)$\downarrow$} &  \makecell[c]{SS\\(\%)$\uparrow$} & \makecell{ERS\\(\%)$\uparrow$} & \makecell{WER\\(\%)$\downarrow$} \\
\hline
Origin & 64.0 & 77.7 & 8.0 & 84.2 & 85.6 & 8.4 & 88.9 & 89.8 & 8.4 & 91.2 & 92.0 & {4.0} \\
Full FT & 69.4 & 83.9 & 10.3 & 90.2 & 93.0 & 9.5 & 92.5 & 95.6 & 25.1 & 94.5 & \textbf{97.0} & 12.1 \\
\hline
\multicolumn{13}{l}{\textit{Change the smallest selection}} \\
smallest & \textbf{69.9} & \textbf{83.2} & \textbf{8.5} & \textbf{90.2} & \textbf{94.2} & \textbf{7.0} & \textbf{92.6} & \textbf{95.5} & \textbf{9.2} & \textbf{94.8} & {96.8} & \textbf{3.8} \\
2nd & 69.4 & {83.1} & \textbf{8.5} & 90.0 & 93.8 & 8.0 & 92.3 & 95.0 & 10.3 & 94.4 & 96.0 & 5.0 \\
3rd & 69.4 & 83.0 & 9.8 & 89.5 & 92.7 & 8.9 & 92.1 & 93.9 & 11.9 & 94.0 & 95.2 & 6.4 \\
\hline
\multicolumn{13}{l}{\textit{Change the largest selection}} \\
largest & \textbf{69.9} & \textbf{83.2} & \textbf{8.5} & \textbf{90.2} & \textbf{94.2} & \textbf{7.0} & {92.6} & {95.5} & \textbf{9.2} & \textbf{94.8} & {96.8} & \textbf{3.8} \\
2nd & 68.4 & 82.4 & 11.3 & 89.7 & 93.1 & 10.0 & \textbf{92.6} & \textbf{95.7} & \textbf{9.2} & 94.2 & 95.8 & 6.6 \\
3rd & 68.2 & 82.2 & 10.3 & 89.4 & 92.5 & 9.7 & 92.5 & 95.4 & 14.4 & 93.8 & 95.0 & 8.2 \\
\hline
\end{tabular}
}
\label{ablation2}
\end{table}

\section{Conclusion}
This paper presents \our{}, a selective partial fine-tuning strategy guided by layer-wise characteristic-specific analysis. By identifying the varying contributions of Transformer layers to emotion and speaker control through a weighted-sum framework, we selectively adapt only the layers with the highest and lowest importance. 
Experimental results across four large-scale codec language TTS models demonstrate that \our{} matches or exceeds full fine-tuning performance while updating only 8\% of the parameters and accelerating training by 2$\times$. The identified layer importance profiles also exhibit strong cross-dataset transferability, providing an efficient and robust solution for high-quality, resource-saving TTS domain adaptation.

\section{Generative AI Use Disclosure}
During the preparation of this manuscript, the authors used generative AI tools to polish the English language, improve readability, and assist with \LaTeX{} formatting. These tools were not used to generate any scientific claims, experimental results, or significant parts of the manuscript.

\bibliographystyle{IEEEtran}
\bibliography{mybib}

\end{document}